\documentclass[pre,aps,preprint,showpacs]{revtex4}
\usepackage{graphicx,subfigure}

\begin{document}
\author{Marian Brandau and Steffen Trimper}
\affiliation{Fachbereich Physik, Martin--Luther--Universit\"at,D--06099 Halle Germany}
\email{trimper@physik.uni-halle.de}
\title{Network of social groups or Let's have a party}
\date{\today }
\begin{abstract}
We present a simple model for growing up and depletion of parties due to the permanent communication 
between the participants of the events. Because of the rapid exchange of information, everybody is able 
to evaluate its own and and all other parties by means of the list of its friends. Therefore the number of 
participants at different parties can be changed incessantly. Depending on the deepness of the social contacts, 
which will be characterized by a parameter $\alpha $, a stable distribution of party members emerges. At a 
critical $\alpha _c$ an abrupt depletion of almost all parties is observed and as the consequence all the 
peoples are assembled at a single party. The model is based on a hierarchical social network. The probability 
that a certain person is contacted to another one depends on the social distance introduced within the network 
and homophily parameter $\alpha $.
\end{abstract}

\maketitle

\section{Introduction}
Finding of groups of alike elements in data and in interests is of great importance in all quantitative 
sciences. Thereby methods and tools of theoretical physics turned out to be successful in giving new insight into 
the complex behavior in different situations. Models developed in statistical physics proved to be 
fruitful in reproducing and predicting features of traffic \cite{hel}, migration problems \cite{weid} as well 
as in opinion formation within social groups \cite{kac,schw}. Although the modeling of opinion 
formation, based on application of cellular automata, Ising models and other tools of computational physics,  
is a quite drastic simplification of complicated cognitive processes, the general properties of 
the system will be reflected \cite{stau}. The main goal of the paper dealing with opinion formation, for 
a review see \cite{stau}, are interested to figure out a complete consensus or a diversity of final opinions  
from initially diverging opinion. As a further simplification some authors \cite{kac,szna} have used a binary 
choice for opinions. Thus, in the Sznajd model \cite{szna}, two people sharing the same opinion tend to convince its 
(six) neighbors of the pair opinion. A generalization of the model including advertising and aging in a multilayer 
Sznajd model is discussed in \cite{schu2}. Another related topic is the ghetto formation studied by means of the two-dimensional Ising model with Kawasaki-exchange dynamics \cite{mey}. An extension to multi-cultural societies is 
discussed recently in \cite{schu1}. A further enlargement of the model is emerged by allowing many discrete opinions 
on a network \cite{smo,stau2}. Hereby the analysis is based on a directed network due to Albert and Barab\'{a}si 
\cite{ab}, for a discussion of networks see also \cite{dorog}.

Our paper is likewise addressed to a relation between networks and dynamics of social groups. In particular, we  
study different individuals within a social group with conflicting interests. Different to other approaches the 
conflicts may also occur during the time evolution and not only initially, e.\,g. physically the interaction is time-dependent. To be specific, let us consider young peoples, planning to participate in a common Saturday party. However they are indecisive at which party 
they will attend. To make a decision they are forced to establish everlasting contacts to the other members of 
the society under consideration. Due to the permanent contacts the actual decision of each member is 
modified permanently. Insofar, we consider a dynamical, time dependent interaction between the members of the group.   
Obviously, the characterized situation corresponds to the reality where the peoples taking part at a certain party 
are often bored by the audience of that party. As a consequence they decide to leave the present party and to 
orient towards another one. The change-over to another party is normally triggered by the number of friends 
attendant at that party. The goal of the present paper is to model such a situation. Specifically, we are interested 
in depletion of all parties with exception of one party, reported in a German news magazine \cite{Spiegel}. 
Such a kind of phase transition is also observed in a class of models when a strong leader and external impact are 
present \cite{kac}. In our model the phase transition is due to self-organization effects and is not controlled by 
external environment. Instead of that we have introduced a hierarchical social network discussed recently 
by Watts et al \cite{watts}. This network captures the essential ingredients of a network model of connected 
population \cite{ab,wast,bar,stro}. The construction of such kind of networks is motivated by the observation that 
individuals within a certain population may be grouped according to their function in the society, for example, their 
hobby, their occupation, etc. The model offers the phenomena of ``six degree of separation'' discovered by Travers 
and Milgram some years ago \cite{tm}. Recently the spreading of epidemics within a hierarchical network has been 
discussed \cite{zhtz}. Here, we analyze a party model on a hierarchical social network under a permanent self-organized 
interaction between the agents.

\section{The Model} 

\subsection{The Society}
Let us characterize the situation we are interested in by a group of predominantly young peoples willing to 
organize some events like parties. However, they are not up to visit any party. Instead of that their aim is 
to find out the ``best'' party at the evening. To decide what the best party is, the peoples communicate 
permanently with their friends getting stuck at any party. Due to everlasting exchange of opinions  
by mobile phone everyone is well informed about the status of any party. Each member of the social 
group standing about such an event is able to evaluate its own party by comparison with other ones. Based on the 
permanent ability getting the total information, everybody decides on its stay at the present party or to change 
the party. The decision is strongly influenced by the behavior of subgroups which we identify with the circle 
of friends. 
In that sense our model is also a stochastic one because the members of the group make their decision, the 
disposition to change the party stochastically. This kind of emergent behavior is characteristic for a social 
group and hence, there is an evidence to adopt method of statistical mechanics. Hereby, the ensemble is given by 
a certain group the members of which are linked by common interests such as to arrange parties. They form a 
social network within the society. The permanent exchange of information and the subsequent decision to change 
the party or to stay at the party is related to the interaction between the constituents. Thereby, the distribution 
of parties with different attractiveness offers an additional interaction for the members of the social group under consideration. To be more specific, let us assume the group consists of $N$ persons where all the friends should 
be included. Further, the systems contains $V$ nodes and $E \subseteq V \times V$ edges. The network is 
characterized by the numbers $(V,E)$. The nodes $x^{i} \in V$ represents the agents or the persons of the game,  
whereas the edges between the nodes stands for the connections or the acquaintances between the persons. 
Additionally, each node $x^i$ is assigned to a set of neighbors $y_{\gamma}^{i}$ with $i \in [1,N]$ and 
$\gamma =1,\ldots,k$. This set is called the friends. The single adjacency list, denoted by $A_i \rightarrow (y_{1}^{i},\dots,y_{k}^{i})$, consists of all the pairs $(x^i,y_1^i), (x^i,y_2^i), \ldots, (x^i,y_k^i) \in  E$ with 
$i \in [1,N]$. The number $k$ is identified as the degree of the set $x^i$. Each adjacency list $A=\{A_i\}$ presents one realization of the network.
\subsection  {Hierarchical Networks}
\begin{figure}
\centering
\includegraphics[width=10cm]{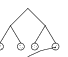}
\vspace{1cm}
\caption{Schematic diagram of grouping individuals in a hierarchical social network with $\ell = 3$. Each group is further divided into $b = 2$ subgroups. A group of $N$ nodes 
are classified into lowest-level subgroups with $\langle g \rangle  = 5$. There $x_{ij}$ is the 
social distance between nodes $i$ and $j$ and is here $x_{ij} = 3$.}   
\label{Fig.1}
\end{figure}

Following the basic idea by Watts et al \cite{watts} let us introduce a hierarchical network. The model was motivated 
by the general structure in the groupings of individuals in a society. Such a classification reflects the deepness 
of relations for instance by the families, the working team, the hobbies or the home district. The situation is 
depicted in Fig.~\ref{Fig.1} schematically \cite{stro}. The highest level can be regarded as a population of $N$ 
individuals or nodes. This $N$ nodes may then be partitioned into $b$ groups, each of them can further divided into $b$ subgroups and so on. 
After $(\ell-1)$ divisions the structure has a total of $\ell$ levels. The underlying structure 
ends at a level where an individual belongs to a close functional group of size $g$ where $1\leq g \leq N $ is 
typically of the order $10^1$ to $10^2$. The members belonging to the lowest-level subgroup have the highest change 
of getting friends or in general becoming similar. As stressed in \cite{zhtz} the division in subgroups is usually 
not unique. For example, all physicists in a certain university can be classified roughly by their research area but 
simultaneously they can be grouped geographically based on the region where their institute is located. As usually the 
number of persons taking part in a party (nodes) may thus be characterized by $H$ hierarchies. Each of them takes on 
the structure shown in Fig.~\ref{Fig.1}. As a further important quantity within a hierarchy let us introduce the social 
distance $x_{ij}$ that measures the similarity between two nodes $i$ and $j$. In case of different hierarchies we define 
$x_{ij} = \max_H x_{ij}^{H}$. For nodes belonging to the same lowest-level group in a given hierarchy, 
$x_{ij} = 1$, otherwise the social distance is even the number of levels from the lowest for which the nodes belong to the same group.

The probability that a person $x^i$ is linked to a person $x^j$, specifying that $i$ and $j$ are friends, is 
established with a probability 
\begin{equation} 
P(x_{ij}) = \frac{\exp(-\alpha x_{ij})}{\sum_{n=1}^\ell \exp(-\alpha n)} \label{eq1}
\end{equation}
Here the parameter $\alpha$ is a measure of homophily of the system. With other words, the quantity $\alpha$ 
characterizes the deepness of the contacts between two nodes (persons) within the system. The probability, 
given in Eq.~(\ref{eq1}), guarantees that for $\alpha \gg 1$ only links between nodes with small 
separation, e.\,g. such belonging to the same subgroup, are probable, whereas for $\alpha = - \ln b$, links between 
individuals with any social distance are equally probable. In that case a random network results. For intermediate 
values of $\alpha$, the network shows small-world features.

The above division process is repeated until a mean number of $\langle z \rangle =  \langle g \rangle - 1$ links are established for each 
individual in the system. Here $\langle z \rangle$ is the averaged number of friends within the network. Concluding this 
section we emphasize the model is characterized by the set of parameters $N$, $H$, $b$, $\ell$, $g$ and $\langle z \rangle$ with the number 
of nodes $N = \langle g \rangle b^{\ell-1}$. The quantity $\langle g \rangle$ is the average size of the lowest-level subgroups.
\section{Party modelling and results}

\subsection{The network} 

\begin{figure}
\centering
\subfigure[]{\label{F2.a} \includegraphics[width=6cm]{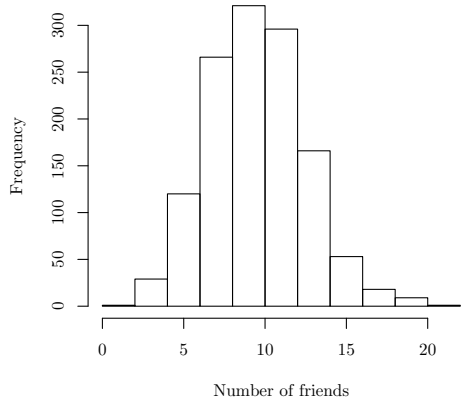}}
\subfigure[]{\label{F2.b} \includegraphics[width=6cm]{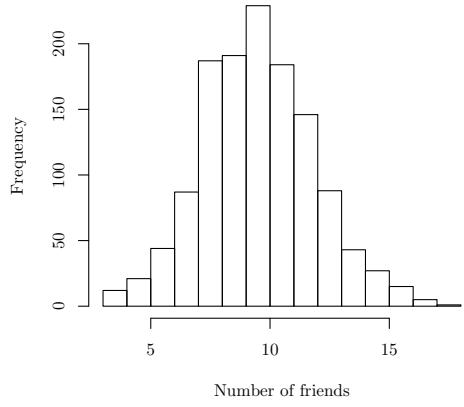}}
\subfigure[]{\label{F2.c} \includegraphics[width=6cm]{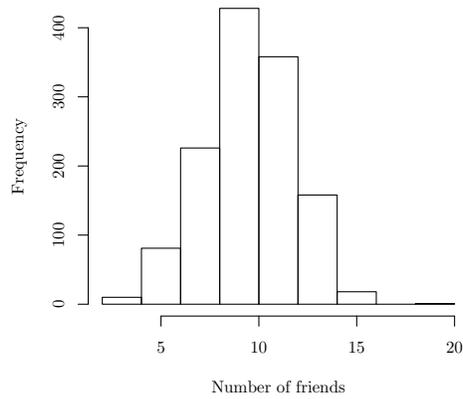}}
\caption{Frequency distribution of neighbors or friends for three different parameters \ref{F2.a} $\alpha = - \ln 2$, \ref{F2.b} $\alpha = 1$ and \ref{F2.c} $\alpha = 10$. 
The number of nodes is $N = 1280$ ($\ell = 8$). The distribution is strongly determined by the network.}
\label{Fig.2}
\end{figure}

A stressed above the social network will be constructed according to \cite{watts}. For a population of $N$ individuals 
the structure is shown in Fig.~\ref{Fig.1}. The links between different nodes (agents, persons) are made by the 
following steps. Firstly we chose a node $x^i$ from the set of all nodes randomly. Then we take a link of the distance 
$x_{ij}$ with the probability given by Eq.~(\ref{eq1}). In a next step we select from all nodes with the given 
distance $x_{ij}$ from $x^i$ a second node $x^j$. After this a link between the nodes $x^i$ and $x^j$ is established. 
Such a link specifies that $i$ and $j$ are friends, e.\,g. more mathematically, the node $x^j$ is added to the adjacency 
list $A_i$ and the node $x^i$ is added to $A_j$. This procedure is repeated until the persons in the network 
exhibit a mean number $\langle z \rangle$ of friends. As the result we obtain a set of adjacency lists $A$ which characterizes 
the network totally. In Fig.~\ref{Fig.2} we show the frequency distribution of neighbors (or friends or contacts) for different values of the homophily parameter $\alpha$. Let us point out that the distribution is determined by the network
essentially. The results have to be incorporated in the further consideration.

To initialize the parties into the model there exits different possibilities. Once one could chose a special distribution 
of the parties initially, or there is a preselection in such a manner that some places for parties are favored in 
advance. This situation is not considered here, because we are interested in the self-organization mechanism. While the parties are in progress, the participants want to come to a decision. Instead of that we apply a random distribution of 
parties.

In the next step we discuss the procedure for making the decision to leave or to stay at a party. For illustration 
let us assume that a person $i$ is on a certain party denoted as party number 5. The friend of $i$, labeled by $j$, 
is on another party, for instance that one with number 2. As a result of the phone call of $i$ with $j$  
there appears two possibilities for $j$. Either $j$ remains on its party or $j$ changes from its party 2 to party 5. 
As a criteria for the decision person $j$ disposes of information of its own party, denoted by $\mathbf{a}$ and of the 
other party by communication, characterized by, lets say $\mathbf{b}$. Now we define a function $(\mathbf{a},\mathbf{b})$, specified below, the result of which is the decision ``go'', in case the party of $i$ is better than the party of its own party 
($j$'s party) and is ``stay'' in the opposite case. In our realization person $x^j$ knows the number of friends on its 
party. Thus let us chose $\mathbf{a} = \#(A_j \cap M_j)$, where $M_j$ is total number of peoples at the party $j$. 
Likewise the person $x^j$ knows the number of friends on $i$'s party. Consequently we chose $\mathbf{b} = \#(A_j \cap M_i)$, where $M_i$ is the total number of peoples at the party $i$. A very simple, but realistic rule, is to take the decision $(\mathbf{a},\mathbf{b}) \rightarrow $ ``go'', whenever $\mathbf{a} < \mathbf{b}$. In the opposite case $\mathbf{a} \geq \mathbf{b}$ the decision is ``stay''. In case the number of friends of the other party is larger than on the own party the person decides to change. 

\subsection{Results} 
\begin{figure}
\centering
\includegraphics[width=12cm]{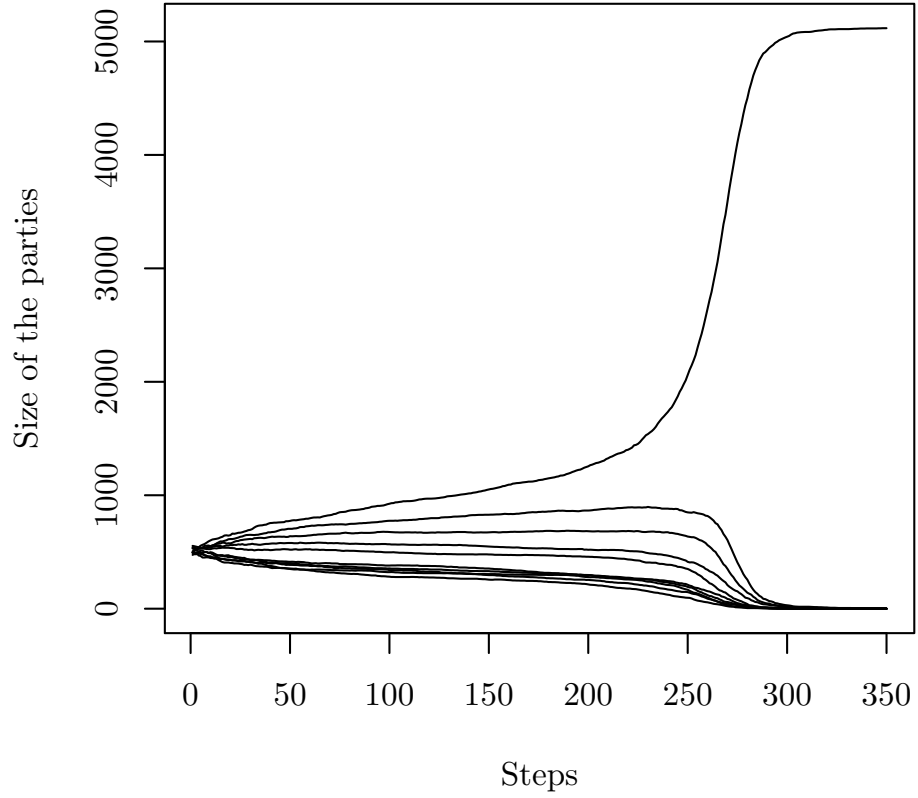}
\caption{Long-time evolution of the size of the parties for $\alpha = -\ln b$, $b =2$, $H = 1$ and $N = 5120$. The number of 
parties is 10.}
\label{Fig.3}
\end{figure}

\begin{figure}
\centering
\includegraphics[width=12cm]{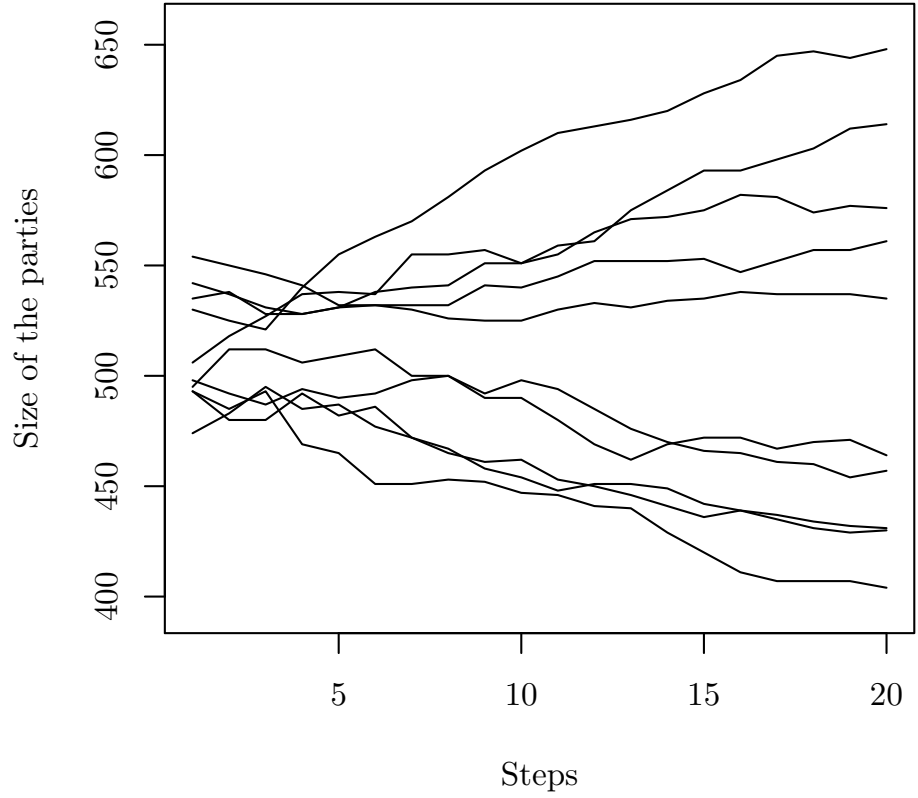}
\caption{Short-time expansion of the size of parties for the same parameters as in Fig.~\ref{Fig.3}. Obviously, the biggest party 
initially will not be the winner in the long-time expansion.}
\label{Fig.4}
\end{figure}

\begin{figure}
\centering
\includegraphics[width=12cm]{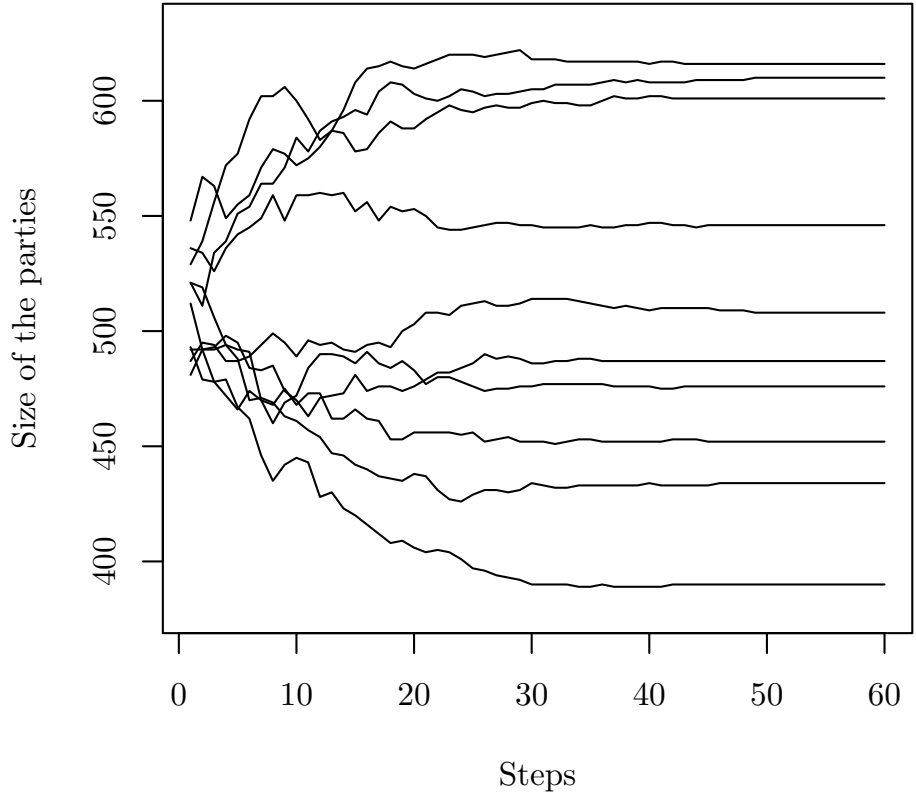}
\caption{Long-time expansion of the size of the parties for large parameter $\alpha = 10$. In that case only 
links between nodes with small separation $x_{ij}$ are probable. Large values for $\alpha$ leads to isolated 
subgroups of nodes. }
\label{Fig.5}
\end{figure}

\begin{figure}
\centering
\subfigure[]{\label{F6.a} \includegraphics[width=6cm]{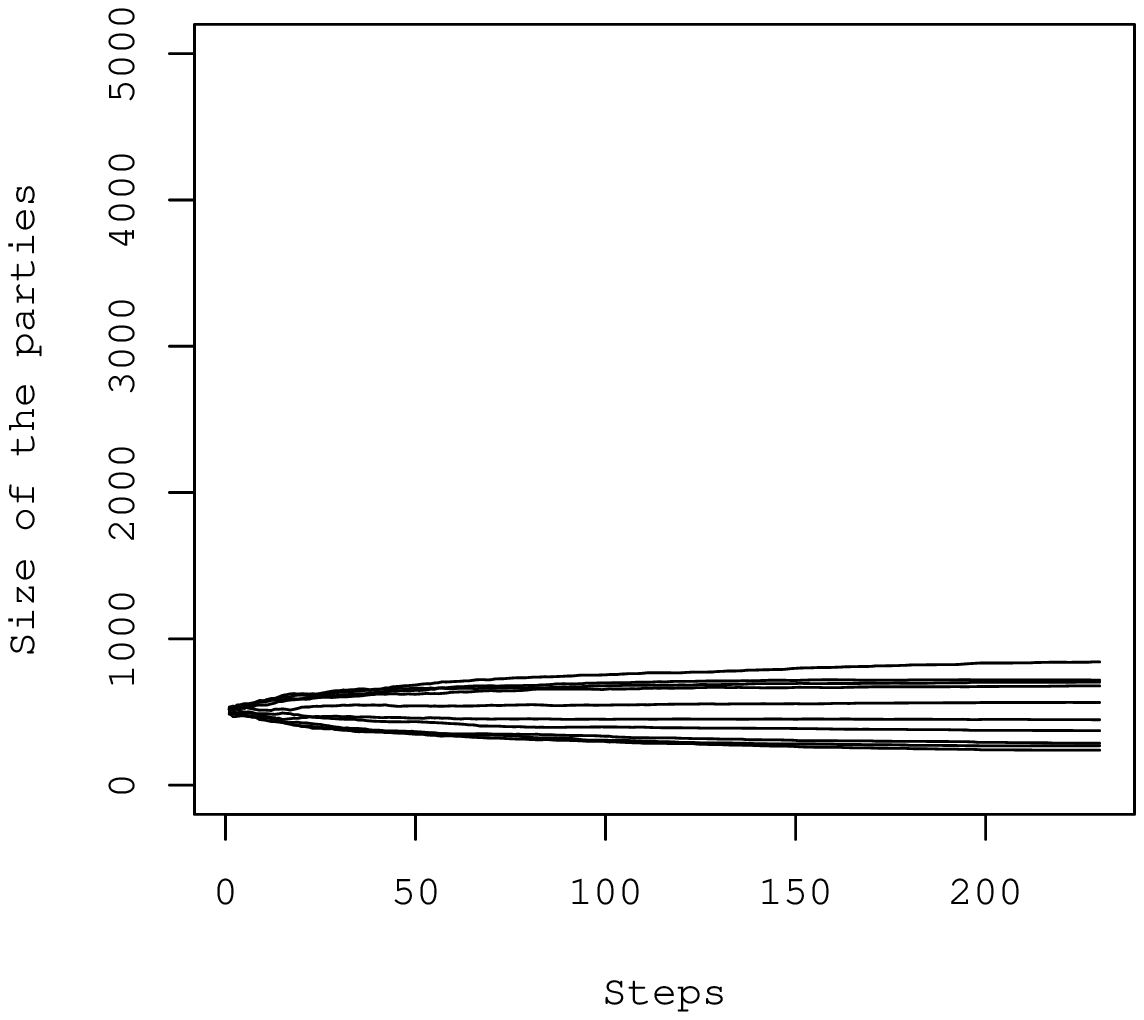}}
\subfigure[]{\label{F6.b} \includegraphics[width=6cm]{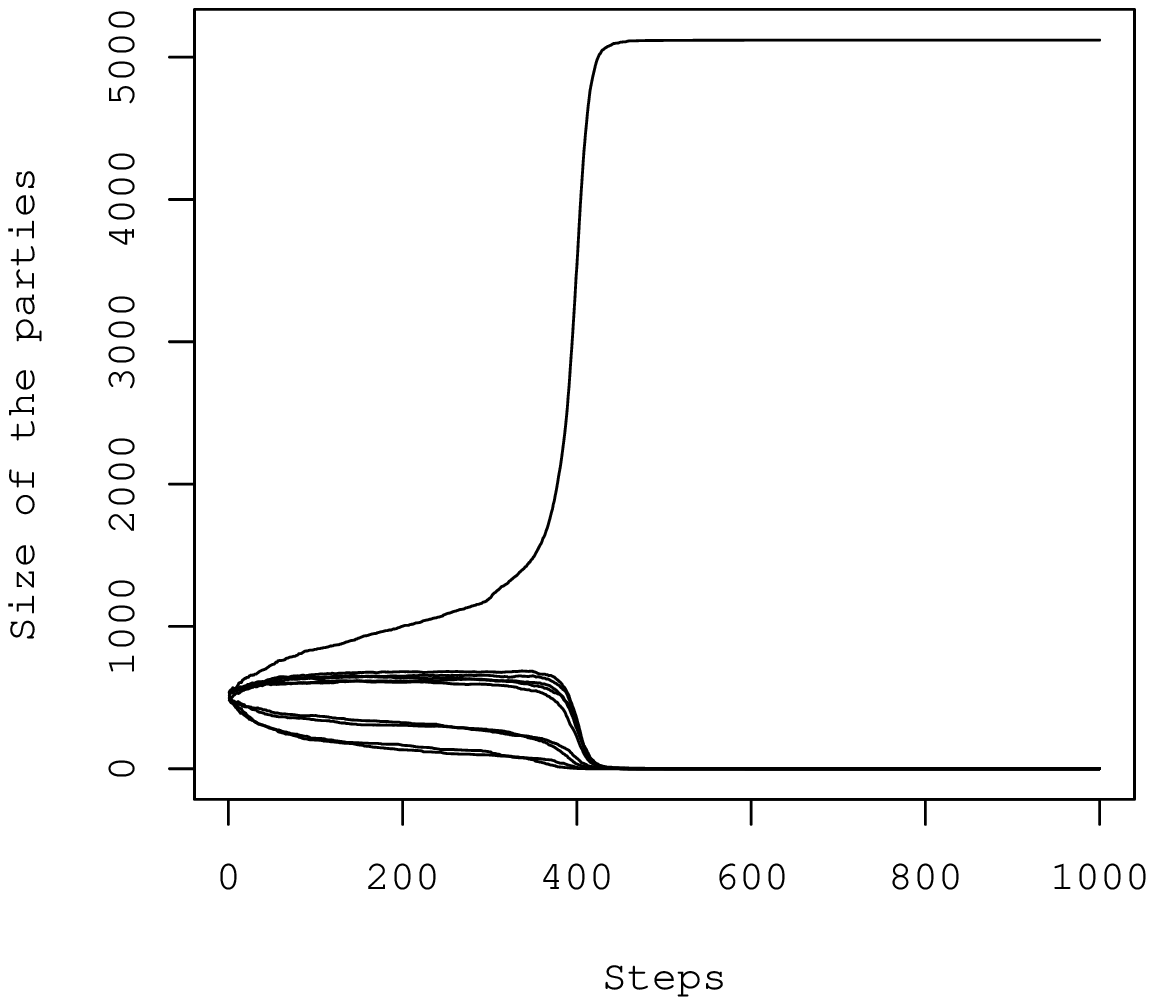}}
\caption{Behavior of the size of parties in the vicinity of the critical value $\alpha _c = - 0. 198$. Whereas for 
\ref{F6.b} $\alpha > \alpha _c$ only one isolated party survives a many party state is observed for \ref{F6.a} $\alpha < \alpha _c$.}
\label{Fig.6}
\end{figure}

In this part the results of the simulation are presented. In particular, we want to demonstrate the essential influence of the parameter $\alpha$ introduced in Eq.~(\ref{eq1}). In Fig.~\ref{Fig.3} the long-time expansion for the size of parties is shown for $\alpha = - \ln 2$. One observes that one party is the winner of the competition. All other parties 
deplete during the course of the evening. As depicted in Fig.~\ref{Fig.4} this situation is not preprogrammed. 
In that figure the short-time expansion of the party-size is shown for the same parameter $\alpha = -\ln 2$ as in 
Fig.~\ref{Fig.3}. Not that party will be best one at which initially the most persons have been present. Insofar 
our model seems to be ergodic. However, this point needs further studies. A very typical situation is offered in 
Fig.~\ref{Fig.5}. Here the homophily parameter is large $\alpha \gg -\ln b$. As stressed in the last section
this guarantees that only links between nodes with small separation are probable. The behavior of the system will be 
dominated by small or isolated subgroups. The same situation is also observed for other intermediate values of $\alpha$. 
Now let us consider the case that only party ``survives'' with higher accuracy. Such a situation is already shown in 
Fig.~\ref{Fig.3}. In Fig.~\ref{Fig.6} we show the results for the region $- \ln 2 \leq \alpha \leq 1$. For $\alpha = \alpha _c \equiv -0.198$ the system undergoes a phase transition from a single party state for 
$\alpha = -0. 2 > \alpha _c$ to a multi-party state for $ \alpha = -0.197 < \alpha _c$. The sharp increase 
is not due an external impact but exclusively by the internal, self-organized interaction between the members of 
the social group. As demonstrated by a slight modification of the parameter $\alpha$ near to $\alpha _c$ the result 
is stable. A similar phenomenon but in another context was observed in \cite{ChOp}.

\section{Conclusions}
In the present paper we have established a simple model to study the behavior of (young) peoples taking part at 
different parties within a large city. Due to the mobile phones they are able to exchange information permanently. 
This fact enables the group an everlasting evaluation of the respective party at which they are present. As a 
simple but realistic measure of the deepness of contacts we have introduce a list of friends labeled to each person. 
Governed by the aim, to be at the ``best'' party, the number of friends at a certain party is casting for a decision to 
leave or to stay at the party. In according to a majority rule any person decides spontaneously to leave its party. An important role for the social contacts between the involved persons is played by the homophily parameter $\alpha$ introduced in Eq.~(\ref{eq1}). Depending on the value of $\alpha$ we observe different scenarios. There is a critical value $\alpha _c$ at which the system offers a phase transition from a multi-party behavior to a single-party state. It would be interesting to study an analytical approach based on a $q$-state Potts-model as suggested in \cite{rb} for fuzzy community structures or \cite{tak} in case of financial market simulations.

\begin{acknowledgments} 
We acknowledge discussions with Gunter Sch\"{u}tz (Fz. J\"{u}lich) and collaboration with Dafang Zheng (Hangzhou). 
The paper had been supported by the DFG under the grant TR 3000/3--3.
\end{acknowledgments}

\newpage

\newpage
\begin{thebibliography}{90}
\bibitem{hel}D. Helbing, Rev. Mod. Phys. {\bf 73}, 1067 (2001). 
\bibitem{weid}W. Weidlich, G. Haag, {\em Concepts and Models of Quantitatively Sociology} (Springer, Berlin 1983); 
W. Weidlich, Phys. Rep. {\bf 204}, 1 (1991)
\bibitem{kac}K. Kacperski and J. A. Holyst Physica A {\bf 269}, 511 (1999); ibid {\bf 287}, 631 (2000); 
\bibitem{schw}F. Schweitzer and J. A. Holyst, Eur. J. Phys. B {\bf 15}, 723 (2000).  
\bibitem{stau}D. Stauffer {\em How to convince others ?} in AIP Conference on the Monte Carlo method in the physical sciences edited by J. E. Gubermatis (2003), cond-mat/0307133.   
\bibitem{szna} K. Sznajd-Weron and J. Sznajd, Int. J. Mod. Phys. C {\bf 11}, 1157 (2000).
\bibitem{schu2}C. Schulze, Int. J. Mod. Phys. C {\bf 14}, 95 (2003).
\bibitem{mey} H. Meyer-Ortmanns, Int. J. Mod. Phys. C {\bf 14}, 311 (2003).
\bibitem{schu1} C. Schulze, Int. J. Mod. Phys. C {\bf 16}, 351 (2005).
\bibitem{smo} D. Stauffer and H. Meyer-Ortmanns, Int. J. Mod. Phys. C {\bf 15}, 241 (2004).
\bibitem{stau2}D. Stauffer, A. O. Sousa, and C. Schulze {\em Discretized opinion dynamics of Deffuant model on scale-free networks}, cond-mat/0310243 (2004). 
\bibitem {ab} R. Albert and A.-{L}. Barab\'{a}si, Rev. Mod. Phys.{\bf 74}, 47 (2002).
\bibitem {dorog} S. N. Dorogovtsev and J. F. F. Mendes, {\em Evolution of Networks} (Oxford University Press, 
New York 2003). 
\bibitem{Spiegel} {\em Keine Ahnung $\dots$ ich ruf dich an}, Der Spiegel {\bf 12} (2004).
\bibitem{watts} D. J. Watts, P.S. Dodds, and M. E. J. Newman, Science {\bf 296}, 1302 (2002).
\bibitem{wast} D. J. Watts and S. H. Strogatz, Nature {\bf 393} 440 (1998).
\bibitem {bar}A.-{L}. Barab\'{a}si and R. Albert, Science {\bf 286}, 509 (1999).
\bibitem{stro}S. H. Strogatz, Nature {\bf 410}, 268 (2001).
\bibitem {tm}J. Travers and S. Milgram, Sociometry {\bf 32}, 425 (1969).
\bibitem{zhtz}D. Zheng, P. M. Hui, S. Trimper and B. Zheng, Physica A {\bf 352}, 659 (2005).
\bibitem{rb}J. Reichardt and S. Bornholdt, Phys. Rev. Lett. {\bf 93}, 218701 (2004).
\bibitem{tak}T. Takaishi, {Simulations of financial makets in a Potts-like model}, cond-mat/0503156.
\bibitem{ChOp} A. Pluchino, V. Latora and A. Rapisarda, \textit{Int. J. Mod. Phys. C} \textbf{16}, 515 (2005)
\end{thebibliography}
\end{document}